\documentclass{article}
\usepackage{spconf,amsmath,graphicx}
\PassOptionsToPackage{hyphens}{url}\usepackage{hyperref}

\usepackage{here}
\usepackage{amsmath}
\usepackage{amsmath,bm,amssymb,url,cite}
\usepackage{multirow}
\usepackage{xcolor}
\usepackage{cite}

\usepackage{makecell}
\usepackage{multicol}
\usepackage{tabularx}

\def\tthline{\noalign{\hrule height 1.4pt}}

\newcommand\tabref[1]{Table~\ref{#1}}
\newcommand\secref[1]{Section~\ref{#1}}

\newcommand{\redit}[1]{\textcolor{black}{#1}}
\newcommand{\kedit}[1]{\textcolor{black}{#1}}
\newcommand{\sedit}[1]{\textcolor{black}{#1}}
\newcommand{\rredit}[1]{\textcolor{black}{#1}}
\newcommand{\kkedit}[1]{\textcolor{black}{#1}}
\newcommand{\ssedit}[1]{\textcolor{black}{#1}}
\newcommand{\rrredit}[1]{\textcolor{black}{#1}}

\title{\kkedit{Lightweight} and high-fidelity End-to-end text-to-speech\\
with multi-band generation and inverse short-time Fourier transform
}

\name{Masaya Kawamura$^{1*}$\thanks{*Work performed during an internship at LINE Corporation.}, Yuma Shirahata$^{2}$, Ryuichi Yamamoto$^{2}$, and Kentaro Tachibana$^{2}$}
\address{$^{1}$The University of Tokyo, Japan, $^{2}$LINE Corp., Japan.}
\begin{document}

\fontsize{9.5}{11.3}\selectfont
\maketitle
\begin{abstract}
We propose a lightweight end-to-end text-to-speech model using multi-band generation and inverse short-time Fourier transform.
Our model is based on VITS, a high-quality end-to-end text-to-speech model, but adopts two changes for more efficient inference: 1) the most computationally expensive component is partially replaced with a simple inverse short-time Fourier transform, and 2) multi-band generation, with fixed or trainable synthesis filters, is used to generate waveforms.
Unlike conventional lightweight models, which employ optimization or knowledge distillation separately to train two cascaded components, our method enjoys the full benefits of end-to-end optimization. 
Experimental results show that our model synthesized speech as natural as that synthesized by VITS, while achieving a real-time factor of 0.066 on an Intel Core i7 CPU, 4.1~times faster than VITS. Moreover, a smaller version of the model significantly outperformed a lightweight baseline model with respect to both naturalness and inference speed. 
\rredit{Code and audio samples are available from \url{https://github.com/MasayaKawamura/MB-iSTFT-VITS}.}

\end{abstract}
\vspace{-1mm}
\vspace{-2mm}
\section{Introduction}
\vspace{-1mm}
Text-to-speech (TTS) has undergone a significant improvement thanks to the development of deep learning TTS models~\cite{ze2013statistical,ning2019review,tan2021survey}. 
However, because recent TTS models typically employ large neural networks with millions of parameters to synthesize high-fidelity speech, their slow inference speed has become problematic \rredit{in real-world applications where computational resources are often limited.}
Hence, enabling fast TTS inference under such conditions while preserving synthesis quality is currently one of the most active research topics in TTS~\cite{tan2021survey,kalchbrenner2018efficient,valin2019lpcnet,luo2021lightspeech,zhai2020squeezewave}.

In this context, many approaches have been proposed during the past few years to improve both acoustic models, which generate acoustic features from text~\cite{huang2020devicetts,luo2021lightspeech}, and vocoders, which synthesize waveforms from the predicted features~\cite{kalchbrenner2018efficient,valin2019lpcnet,yang2021mbmelgan,multistream,zhai2020squeezewave}. 
Although these methods successfully speed up the TTS inference even with limited computational resources, the separate optimization of two dependent models (i.e., acoustic models and vocoders) inherently restricts the performance of TTS systems.

To overcome the performance restriction, some previous studies have performed end-to-end optimization of lightweight TTS models~\cite{nix-tts, nguyen21e_interspeech}. In particular, Nix-TTS~\cite{nix-tts} successfully uses a teacher-student framework to distill the knowledge of a high-quality end-to-end TTS model (namely, VITS~\cite{kim2021conditional}) to a smaller student model, while achieving faster inference speed than LiteTTS~\cite{nguyen21e_interspeech}.
However, although Nix-TTS employs an end-to-end model as a teacher, the student model is still trained with two-stage optimization and thus cannot enjoy the full benefits of end-to-end models.

In this paper, we propose \rredit{a lightweight} end-to-end TTS system capable of performing fast inference while achieving high-fidelity waveform generation \kkedit{for on-device applications.} 
We adopted VITS~\cite{kim2021conditional} as the basis of our TTS model for two reasons: 1) the VITS enables high-quality speech synthesis, and 2) its architecture is fully non-autoregressive, which is desirable for achieving fast inference.

To discover which part of the VITS model should be improved, we first investigated its inference bottleneck. The investigation revealed that the decoder part, which transforms latent acoustic features to waveforms, is the most computationally expensive. Therefore, to focus on speeding up the decoder module of VITS, we first replaced a part of the decoder computation with a simple inverse short-time Fourier transform (iSTFT), inspired by iSTFTNet~\cite{iSTFTnet}, to efficiently perform frequency-to-time domain conversion.
To further speed up the inference, we adopted a novel approach that combines the iSTFT-based sample generation with multi-band processing. Specifically, in the proposed method, each iSTFT module generates sub-band signals, which are subsequently summed to generate the full-band target waveform. This significantly cuts the computational cost and speeds up inference, while maintaining synthesis quality. Moreover, we also investigated an approach to use a trainable synthesis filter for the sub-band signals, inspired by a multi-stream vocoder~\cite{multistream}. 

Experiments demonstrated that the best version of our proposed model retained human-level naturalness as well as VITS does, and achieved a real-time factor (RTF) of 0.066 on an Intel Core i7 CPU, which is 4.1 times faster than the original VITS.
Furthermore, to compare our proposed models with Nix-TTS, we also trained a version of our proposed model that is as small as Nix-TTS. 
Experiments showed that the smaller version of the proposed model could generate speech with a significantly better quality than that generated by Nix-TTS, with a mean opinion score (MOS) of 4.43 (vs. 3.69), while achieving much higher generation speed, with an RTF of 0.028 (vs. 0.062).
\sedit{These results indicate that the proposed method takes advantage of both end-to-end model architecture and speed-up techniques as intended.}

\section{Analysis of VITS}
\subsection{Overview of VITS}
Because the proposed model is constructed upon the end-to-end TTS model VITS~\cite{kim2021conditional}, we briefly introduce this model in this section.

The main generative model of VITS is a variational autoencoder (VAE)\kedit{~\cite{kingma2014autoencoding}} with text-conditional prior distribution. The model is trained to maximize the log-likelihood of waveform $x$ given text $c$. However, because this maximization is intractable, the evidence lower bound (ELBO) is maximized instead:
\begin{equation}
    \log{p_{\theta}}(x|c) \geq {E}_{q_{\theta}(z|x)}\left[\log{p_{\theta}(x|z)}-\log{\frac{q_{\phi}(z|x)}{p_{\theta}(z|c)}}\right], \label{eq:elbo}
\end{equation}
where $z$ is the latent variable of the VAE; $p$ and $q$ \redit{are the true distribution and} approximate posterior distribution\redit{s, respectively;} and $\theta$ and $\phi$ are the model parameters for $p$ and $q$, respectively. 
Because the loss is defined as the negative ELBO, the first term of \eqref{eq:elbo} can be viewed as the reconstruction loss of waveform $x$, given $z$ sampled from the approximate posterior distribution $q_{\phi}(z|x)$. 
The second term is the Kullback-Leibler divergence between the posterior and prior distributions.
During inference, $z$ is sampled from the prior $p_{\theta}(z|c)$ instead of $q_{\phi}(z|x)$, and then fed to the decoder of the VAE to generate the waveform.

The neural networks that model $p_\theta(z|c)$, $q_\phi(z|x)$, and $p_\theta(x|z)$ are called the prior encoder, posterior encoder, and decoder, respectively. We briefly introduce the model architecture of these modules below.

\noindent\textbf{Prior encoder:} The prior encoder predicts the prior distribution from phoneme sequences. It consists of \redit{three} modules: text encoder, duration predictor, and flow~\cite{rezende2015variational}. 
\redit{The} text encoder \kkedit{module} generates a phoneme-level representation using a self-attention-based architecture~\cite{vaswani2017attention}. 
The duration predictor predicts the phoneme durations for inference. 
The \rrredit{target durations for training are} obtained by monotonic alignment search~\cite{kim2020glow}. 
The flow module is used to augment a simple Gaussian prior distribution to a more expressive one. 

\noindent\textbf{Posterior encoder:} The posterior encoder predicts the approximate posterior parameters from linear spectrogram. 
The module is composed of the non-causal WaveNet residual blocks used in Glow-TTS~\kkedit{\cite{oord2016wavenet,kim2020glow}}.
\redit{Note that} this module is not used during inference.

\noindent\textbf{Decoder:} The decoder generates waveforms from the latent variable $z$ sampled from the prior or posterior distribution. The model architecture is based on HiFi-GAN~\cite{kong2020hifi}.

\subsection{Inference speed of each module}\label{sec:bottleneck}
To identify the bottleneck of VITS with respect to inference speed, we analyzed the inference time of some \redit{selected modules of VITS}. 
We calculated the RTF, which is defined as (time taken to synthesize speech) /  (duration of the synthesized speech), as an objective criterion.
The \redit{average} RTF was measured using 100 \redit{sentences} that were randomly selected from the LJ Speech dataset~\cite{ljspeech}.
\redit{As shown in} \tabref{tab: module_result}, the decoder part consumed more than 96~\% of the inference time, and is apparently the largest bottleneck.

\begin{table}[tb] 
\vspace{-2.5mm}
\caption{
\fontsize{9.5}{11.3}\selectfont
\redit{Average RTF for each} module of VITS. \redit{The RTF was measured on a single thread} of an Intel Core i7 CPU@2.7~GHz. Note that some small modules are omitted for simplicity.} 
\vspace{1mm}
\label{tab: module_result}
\centering
\scalebox{0.95}{
\begin{tabular}{cc}\tthline
Module  & RTF \\ \hline
Text encoder & 0.010 \\
Flow & 0.019\\
Decoder & 0.819 \\\hline
Total &  0.849 \\\tthline
\end{tabular}
}
\vspace{-2mm}
\end{table}

\section{PROPOSED METHOD}
\subsection{Motivation and strategy}
\begin{figure*}[t]
\centering 
\includegraphics[scale=0.61]{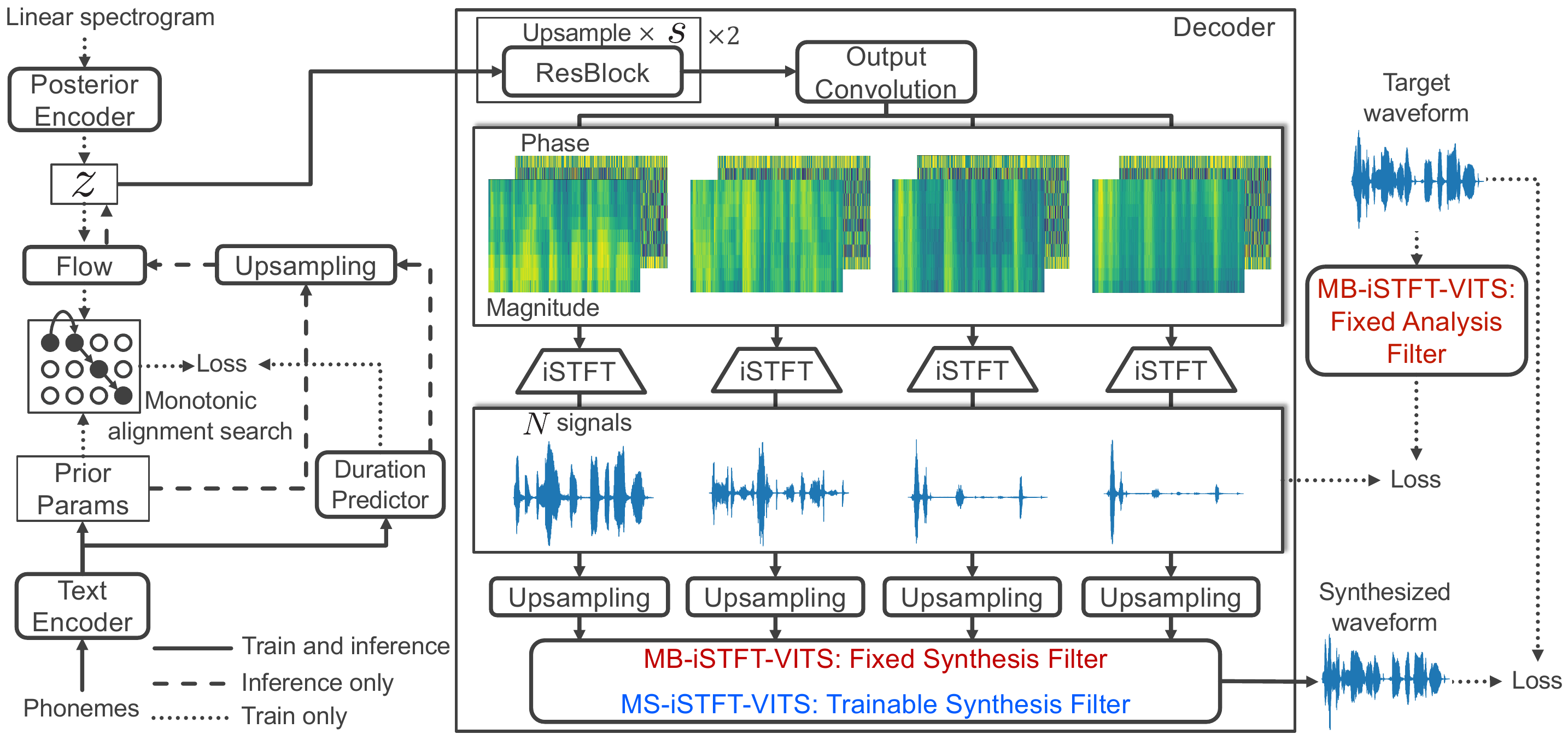}
\vspace{-2mm}
\caption{
\fontsize{9.5}{11.3}\selectfont
Architecture of multi-band iSTFT VITS and multi-stream iSTFT VITS. In multi-band iSTFT VITS, synthesized waveforms are integrated by a fixed synthesis filter. In multi-stream iSTFT VITS, they are integrated by a trainable synthesis filter.}
\label{fig:proposed}
\end{figure*}

The preliminary experiment described in \secref{sec:bottleneck} revealed that the decoder module is the largest bottleneck of VITS. 
Because the decoder architecture is based on the HiFi-GAN vocoder~\cite{kong2020hifi}, which upsamples the input acoustic features with repeated convolution-based network, we first considered reducing the redundancy in this module. 
To this end, we adopted an idea from iSTFTNet~\cite{iSTFTnet}. In this method, some output-side layers of the repeated networks in HiFi-GAN are replaced with simple iSTFT, which significantly reduces the computational cost. 
Specifically, the method is intended to simplify the complex neural vocoding process \rrredit{from} the mel-spectrogram (i.e., performing phase reconstruction and frequency-to-time conversion simultaneously) by replacing the latter with explicit introduction of iSTFT. 
This idea \rrredit{is} also effective in VITS, which performs vocoding from features derived from linear spectrograms.

To further \rredit{improve} the generation speed, we propose an algorithm that combines the iSTFT-based approach with a multi-band parallel strategy, which we introduced in Sections~\ref{sec:multiband} and \ref{sec:multistream}.

\subsection{Multi-band iSTFT VITS}\label{sec:multiband}

There are many studies that have successfully employed a multi-band parallel strategy in the vocoder~\cite{yang2021mbmelgan, yu2019durian, okamoto2018investigation, cui2020efficient}. 
These methods \rredit{exploit} the sparseness of neural networks and use a single shared network to generate all sub-band signals, which significantly cuts the computational cost while maintaining synthesis quality.
Inspired by them, we adopted the same strategy to further \rredit{improve} the inference speed. 

\rredit{Figure~\ref{fig:proposed}} shows the architecture of the proposed model.
As the figure shows, the decoder performs the following processes in a sequential manner.
1) The VAE latent $z$ is upsampled by a factor of $s$ through each convolutional residual block (ResBlock)~\cite{He2016resblock}, where $s$ is a parameter that determines the upsampling scale, and is then projected to the magnitude and phase variables for each of the $N$ sub-band signals. 2) The iSTFT operation is applied to the magnitude and phase variables to generate each sub-band signal. 
3) These sub-band signals are upsampled to match the sampling rate of the original signal by adding zeros between samples, and are then integrated into full-band waveforms using a \rredit{fixed} synthesis filter bank.
Note that the synthesis filter is implemented by a pseudo-quadrature mirror filter bank (pseudo-QMF)~\cite{nguyen1994pqmf}.

During training, the reconstruction loss of VITS is altered to include an additional multi-resolution STFT loss in sub-band scales~\cite{yang2021mbmelgan}.
\redit{To generate \ssedit{the ground truth} sub-band signals, which are necessary to compute the sub-band STFT loss from input waveforms, we use an analysis filter based on pseudo-QMF.}

We refer to this model as multi-band iSTFT VITS (MB-iSTFT-VITS). 
\rredit{Compared with conventional methods that employ separate optimizations for the acoustic model and multi-band vocoder, MB-iSTFT-VITS differs in being optimized in a fully end-to-end manner, and thus achieves better audio quality. }

\subsection{Multi-stream iSTFT VITS}\label{sec:multistream}

Although the multi-band structure enables fast inference with good synthesis quality, the fixed decomposition into sub-band signals can adversely affect the performance of waveform generation because it is an inflexible constraint.
To mitigate this, we also investigated a trainable synthesis filter in the multi-band structure, inspired by \redit{the multi-stream vocoder}~\cite{multistream}. 
This allows the model to decompose speech waveforms in a data-driven manner, which is expected to improve the quality of synthesized speech. 
We refer to this model as multi-stream iSTFT-VITS (MS-iSTFT-VITS). 
Unlike MB-iSTFT-VITS, MS-iSTFT-VITS does not adopt \rredit{a} sub-band STFT loss because the decomposed waveforms in MS-iSTFT-VITS are fully trainable and no longer restricted by fixed sub-band signals.

\section{EXPERIMENTS}\label{sec:eval}
\subsection{Experimental \rredit{conditions}}
We conducted an experiment to evaluate the effectiveness of the proposed method.
We used the LJ Speech dataset\cite{ljspeech} to train and evaluate the models.
This dataset consists of 13,100 short audio clips from a single \rrredit{female} speaker. The total length is approximately 24 hours, and each audio clip is a 16-bit PCM WAV file with a sampling rate of 22,050~Hz. We randomly divided the dataset into a training set (12,500), validation set (100), and test set (500).

\rredit{
We prepared the following five VITS-based TTS models. Note that we adopted a deterministic duration predictor in all models instead of a stochastic one because we found that it stabilizes the prediction of phoneme durations.
}

\noindent\textbf{VITS:}
\rredit{A vanilla VITS of} the official implementation\footnote{ \url{https://github.com/jaywalnut310/vits}}, with the same hyperparameters as the original VITS\cite{kim2021conditional}.

\noindent\textbf{Nix-TTS:}
\rredit{A pretrained model of Nix-TTS\footnote{\url{https://github.com/rendchevi/nix-tts}}.}
The model used was the optimized ONNX version~\cite{bai2019onnx}. 
Note that the dataset used in our experiments is exactly the same as that used for Nix-TTS.

\noindent\textbf{iSTFT-VITS:}
\rredit{A model that incorporates iSTFTNet into the decoder part of VITS.}
The architecture of iSTFTNet is V1-C8C8I, which is the best-balanced model described in \cite{iSTFTnet}. This architecture contains two residual blocks with an upsampling scale of [8, 8]. 
The \rredit{size of fast Fourier transform (FFT)}, hop length, and window length of the iSTFT component were set to 16, 4, and 16, respectively.

\noindent\textbf{MB-iSTFT-VITS:}
\rredit{A proposed model introduced in \secref{sec:multiband}.}
The number of sub-bands $N$ was set to 4. 
The upsampling scale of the two residual blocks was [4,4], to match the resolution of each sub-band signal decomposed by the analysis filters of pseudo-QMF. 
The FFT size, hop length, and window length of the iSTFT component were the same as those used in iSTFT-VITS. Following~\cite{yang2021mbmelgan}, we chose the finite impulse response analysis/synthesis order of 63. To calculate the multi-resolution STFT loss in sub-band scales, the FFT sizes were set to 683, 384, and 171, window lengths to 300, 150, and 60, and hop lengths to 60, 30, and 10, respectively, and a Hann window was used.

\noindent\textbf{MS-iSTFT-VITS:}
\rredit{Another proposed model introduced in \secref{sec:multistream}.}
Following~\cite{multistream}, the kernel size of the convolution-based trainable synthesis filter was set to 63 without bias. 
The other conditions were the same as those of MB-iSTFT-VITS. 

To investigate the performance of the proposed models with a much smaller number of parameters,
we trained a smaller version of MB-iSTFT-VITS\footnote{
\redit{
We selected MB-iSTFT-VITS as the base model architecture for the following reasons: 1) \rrredit{MB-iSTFT-VITS and MS-iSTFT-VITS} performed equally well in terms of naturalness and inference speed, 
and 2) MB-iSTFT-VITS was faster than MS-iSTFT-VITS, with a reduction of $16$ hours in training time.}
}, named Mini-MB-iSTFT-VITS.
We also trained smaller versions of VITS and iSTFT-VITS, named Mini-VITS and Mini-iSTFT-VITS, for comparison.
To construct Mini-MB-iSTFT-VITS, Mini-VITS, and Mini-iSTFT-VITS, 
we halved the number of hidden channels in the text encoder, posterior encoder, flow, and duration predictor. 
In addition, we halved the initial size of the hidden channels in the decoder and the number of layers in the text encoder. As a result, the total number of model parameters was reduced to 7.21~M, which is 3.8 times smaller than the number in the original MB-iSTFT-VITS model. 

For preprocessing, linear spectrograms were obtained from the waveforms by STFT and used as input to the posterior encoder. The FFT size, window size, and hop length were set to 1024, 1024, and 256, respectively.

We used an NVIDIA A100 GPU to train \rredit{all the models except the pretrained model of Nix-TTS}.
The batch size was set to 64 and the models were trained for 800~K steps. We used the AdamW optimizer~\cite{loshchilov2017decoupled} with $\beta_1=0.8, \beta_2=0.99$ and weight decay $\lambda=0.01$. 
The learning rate decay was scheduled by a factor of $0.999^{1/8}$ in every epoch. 
The initial learning rate was set to $2\times10^{-4}$. 

\rredit{To evaluate the quality of the generated speech}, we used a five-point naturalness MOS \kkedit{test}.
Thirty randomly sampled utterances from the test set were evaluated by 18 listeners. 
In addition, we used the number of parameters and RTF as objective criteria.
We measured the average RTF on an Intel Core i7@2.7~GHz using 100 randomly-sampled utterances from the test set.
All models were converted to the ONNX version for evaluation to match the condition to that used in Nix-TTS.
Audio samples are available on our demo page\footnote{\url{https://masayakawamura.github.io/Demo_MB-iSTFT-VITS/}}.

\subsection{Results}
\begin{table}[tb] 
\vspace{-2.5mm}
\caption{
\fontsize{9.5}{11.3}\selectfont
Comparison of model size, naturalness MOS with 95\% confidence intervals, and \rredit{average RTF} on an Intel Core i7@2.7~GHz\kkedit{.}} 
\vspace{1mm}
\label{tab: result1}
\scalebox{0.95}{
\begin{tabular}{cccc}\tthline
Model  &\# Params& MOS & RTF \\ \hline
Ground truth & - &$4.74\pm{0.05}$& - \\
VITS & $28.11$~M& $4.75\pm{0.06}$& $0.27$ \\
iSTFT-VITS& $27.44$~M & $4.65\pm{0.06}$& $0.15$ \\
MB-iSTFT-VITS& $27.49$~M& $4.67\pm{0.06}$ & $0.078$ \\
MS-iSTFT-VITS& $27.49$~M& $4.73\pm{0.06}$ & $0.066$ \\
\Xhline{2\arrayrulewidth}
Nix-TTS & $5.23$~M& $3.69\pm{0.11}$ & $0.062$ \\
Mini-VITS& $7.35$~M& $4.60\pm{0.07}$ & $0.099$ \\
Mini-iSTFT-VITS& $7.19$~M &$4.58\pm{0.06}$& $0.054$ \\
Mini-MB-iSTFT-VITS& $7.21$~M & $4.43\pm{0.08}$& $0.028$ \\\tthline
\end{tabular}
}
\vspace{-2mm}
\end{table}

Table~\ref{tab: result1} shows the experimental results.  
The \rrredit{results showed} that MB-iSTFT-VITS and MS-iSTFT-VITS achieved much faster inference than VITS (with a speedup of 3.4-4.1x), while maintaining \rredit{high} naturalness: the MOS is comparable to that of VITS and the ground truth (no statistically significant difference in student's $t$-test with a 5~\% significance level). 
\rredit{
Although the iSTFT-based approach itself was effective (with a speedup of 1.8x), the performance was enhanced dramatically (with a speedup of 1.9-2.3x) by combining it with multi-band or multi-stream processing.
These results indicate that the proposed method successfully speeds up the inference of the well-designed end-to-end model VITS, while taking advantage of its powerful generative capability.
}

The smaller \kkedit{\textit{mini}} models showed a similar tendency to that of the normal models.
Specifically, we observe that the inference speed of Mini-VITS was less than that of Nix-TTS, which means that simply reducing the number of parameters of the original VITS was not sufficient.
Conversely, the proposed Mini-MB-iSTFT-VITS outperformed Nix-TTS with respect to both inference speed (with a speedup of 2.2x over Nix-TTS) and naturalness (with a MOS increase of 0.74).
This \rrredit{showed} that the smaller version of the proposed model enables both high-speed and high-fidelity waveform generation at the same time, which is desirable \kedit{for on-device applications.} 

Interestingly, there was a trade-off between the inference speed and naturalness for the \kkedit{\textit{mini}} models (Mini-iSTFT-VITS vs. Mini-MB-iSTFT-VITS), which was not observed for normal models (iSTFT-VITS vs. MB-iSTFT-VITS). 
\redit{
We hypothesize that Mini-MB-iSTFT-VITS failed to accurately estimate sub-band signals because of its much smaller number of parameters, and artifacts were caused by imperfect reconstruction of target waveforms.
}

\section{CONCLUSION}\label{sec:conclusion}
In this paper, we proposed an end-to-end TTS system that is capable of high-speed speech synthesis \kedit{for on-device practical applications.} 
Our proposed method is constructed upon a successful end-to-end model named VITS, but employs several techniques to speed up inference, such as reducing the redundancy of decoder computation by iSTFT and adopting a multi-band parallel strategy. Because the proposed model is optimized in a fully end-to-end manner, it enjoys the full benefits of its powerful optimization process, in contrast to conventional two-staged approaches. Experimental results demonstrated that the proposed method can generate speech as natural as that synthesized by VITS, while enabling much faster waveform generation. 
Future research includes extending the proposed method to multi-speaker models.

\vfill\pagebreak

\bibliographystyle{IEEE}
{
\bibliography{strings,refs}

\begin{thebibliography}{10}

\bibitem{ze2013statistical}
{H. Zen}, {A. Senior}, and {M. Schuster},
\newblock ``Statistical parametric speech synthesis using deep neural
  networks,''
\newblock in {\em Proc. ICASSP}, 2013, pp. 7962--7966.

\bibitem{ning2019review}
{Y. Ning}, {S. He}, {Z. Wu}, {C. Xing}, and {L.-J. Zhang},
\newblock ``A review of deep learning based speech synthesis,''
\newblock {\em Appl. Sci.}, vol. 9, no. 19, 2019.

\bibitem{tan2021survey}
{X. Tan}, {T. Qin}, {F. Soong}, and {T.-Y. Liu},
\newblock ``A survey on neural speech synthesis,''
\newblock {\em arXiv preprint arXiv:2106.15561}, 2021.

\bibitem{kalchbrenner2018efficient}
{N. Kalchbrenner}, {E. Elsen}, {K. Simonyan}, {S. Noury}, {N. Casagrande}, {E.
  Lockhart}, et~al.,
\newblock ``Efficient neural audio synthesis,''
\newblock in {\em Proc. ICML}, 2018, pp. 2410--2419.

\bibitem{valin2019lpcnet}
{J.-M. Valin} and {J. Skoglund},
\newblock ``{LPCNet}: Improving neural speech synthesis through linear
  prediction,''
\newblock in {\em Proc. ICASSP}, 2019, pp. 5891--5895.

\bibitem{luo2021lightspeech}
{R. Luo}, {X. Tan}, {R. Wang}, {T. Qin}, {J. Li}, {S. Zhao}, {E. Chen}, and
  {T.-Y. Liu},
\newblock ``Lightspeech: Lightweight and fast text to speech with neural
  architecture search,''
\newblock in {\em Proc. ICASSP}, 2021, pp. 5699--5703.

\bibitem{zhai2020squeezewave}
{B. Zhai}, {T. Gao}, {F. Xue}, {D. Rothchild}, {B. Wu}, {J. E. Gonzalez}, and
  {K. Keutzer},
\newblock ``{SqueezeWave}: Extremely lightweight vocoders for on-device speech
  synthesis,''
\newblock {\em arXiv preprint arXiv:2001.05685}, 2020.

\bibitem{huang2020devicetts}
{Z. Huang}, {H. Li}, and {M. Lei},
\newblock ``{DeviceTTS}: A small-footprint, fast, stable network for on-device
  text-to-speech,''
\newblock {\em arXiv preprint arXiv:2010.15311}, 2020.

\bibitem{yang2021mbmelgan}
{G. Yang}, {S. Yang}, {K. Liu}, {P. Fang}, {W. Chen}, and {L. Xie},
\newblock ``{Multi-band MelGAN}: Faster waveform generation for high-quality
  text-to-speech,''
\newblock in {\em Proc. SLT}, 2021, pp. 492--498.

\bibitem{multistream}
{T. Okamoto}, {T. Toda}, and {H. Kawai},
\newblock ``Multi-stream {HiFi-GAN} with data-driven waveform decomposition,''
\newblock in {\em Proc. ASRU}, 2021, pp. 610--617.

\bibitem{nix-tts}
{R. Chevi}, {R. E. Prasojo}, {A. F. Aji}, {A. Tjandra}, and {S. Sakti},
\newblock ``{Nix-TTS}: Lightweight and end-to-end text-to-speech via
  module-wise distillation,''
\newblock in {\em Proc. SLT}, 2023, pp. 970--976.

\bibitem{nguyen21e_interspeech}
{H.-K. Nguyen}, {K. Jeong}, {S. Um}, {M.-J. Hwang}, {E. Song}, and {H.-G.
  Kang},
\newblock ``{LiteTTS}: A lightweight mel-spectrogram-free text-to-wave
  synthesizer based on generative adversarial networks,''
\newblock in {\em Proc. INTERSPEECH}, 2021, pp. 3595--3599.

\bibitem{kim2021conditional}
{J. Kim}, {J. Kong}, and {J. Son},
\newblock ``Conditional variational autoencoder with adversarial learning for
  end-to-end text-to-speech,''
\newblock in {\em Proc. ICML}, 2021, pp. 5530--5540.

\bibitem{iSTFTnet}
{T. Kaneko}, {K. Tanaka}, {H. Kameoka}, and {S. Seki},
\newblock ``{iSTFTNet}: Fast and lightweight mel-spectrogram vocoder
  incorporating inverse short-time {Fourier} transform,''
\newblock in {\em Proc. ICASSP}, 2022, pp. 6207--6211.

\bibitem{kingma2014autoencoding}
{D. P. Kingma} and {M. Welling},
\newblock ``Auto-encoding variational bayes,''
\newblock in {\em Proc. ICLR}, 2014.

\bibitem{rezende2015variational}
{D. J. Rezende} and {S. Mohamed},
\newblock ``Variational inference with normalizing flows,''
\newblock in {\em Proc. ICML}, 2015, pp. 1530--1538.

\bibitem{vaswani2017attention}
{A. Vaswani}, {N. Shazeer}, {N. Parmar}, {J. Uszkoreit}, {L. Jones}, {A. N.
  Gomez}, {\L. Kaiser}, and {I. Polosukhin},
\newblock ``Attention is all you need,''
\newblock in {\em Proc. NeurIPS}, 2017, vol.~30, pp. 5998--6008.

\bibitem{kim2020glow}
{J. Kim}, {S. Kim}, {J. Kong}, and {S. Yoon},
\newblock ``Glow-tts: A generative flow for text-to-speech via monotonic
  alignment search,''
\newblock in {\em Proc. NeurIPS}, 2020, vol.~33, pp. 8067--8077.

\bibitem{oord2016wavenet}
{A. van den Oord}, {S. Dieleman}, {H. Zen}, {K. Simonyan}, {O. Vinyals}, {A.
  Graves}, et~al.,
\newblock ``{WaveNet}: {A} generative model for raw audio,''
\newblock {\em arXiv preprint arXiv:1609.03499}, 2016.

\bibitem{kong2020hifi}
{J. Kong}, {J. Kim}, and {J. Bae},
\newblock ``{HiFi-GAN}: Generative adversarial networks for efficient and high
  fidelity speech synthesis,''
\newblock in {\em Proc. NeurIPS}, 2020, vol.~33, pp. 17022--17033.

\bibitem{ljspeech}
{K. Ito},
\newblock ``{The LJ speech dataset},''
  \url{https://keithito.com/LJ-Speech-Dataset/}, 2017.

\bibitem{yu2019durian}
{C. Yu}, {H. Lu}, {N. Hu}, {M. Yu}, {C. Weng}, {K. Xu}, et~al.,
\newblock ``{DurIAN}: Duration informed attention network for speech
  synthesis,''
\newblock in {\em Proc. INTERSPEECH}, 2020, pp. 2027--2031.

\bibitem{okamoto2018investigation}
{T. Okamoto}, {K. Tachibana}, {T. Toda}, {Y. Shiga}, and {H. Kawai},
\newblock ``An investigation of subband wavenet vocoder covering entire audible
  frequency range with limited acoustic features,''
\newblock in {\em Proc. ICASSP}, 2018, pp. 5654--5658.

\bibitem{cui2020efficient}
{Y. Cui}, {X. Wang}, {L. He}, and {F. K. Soong},
\newblock ``An efficient subband linear prediction for {LPCNet}-based neural
  synthesis,''
\newblock in {\em Proc. INTERSPEECH}, 2020, pp. 3555--3559.

\bibitem{He2016resblock}
{K. He}, {X. Zhang}, {S. Ren}, and {J. Sun},
\newblock ``Deep residual learning for image recognition,''
\newblock in {\em Proc. CVPR}, June 2016, pp. 770--778.

\bibitem{nguyen1994pqmf}
{T. Q. Nguyen},
\newblock ``Near-perfect-reconstruction pseudo-{QMF} banks,''
\newblock {\em IEEE Trans. Signal Processing}, vol. 42, no. 1, pp. 65--76,
  1994.

\bibitem{bai2019onnx}
{J. Bai}, {F. Lu}, {K. Zhang}, et~al.,
\newblock ``{ONNX}: Open neural network exchange,''
  \url{https://github.com/onnx/onnx}, 2019.

\bibitem{loshchilov2017decoupled}
{I. Loshchilov} and {F. Hutter},
\newblock ``Decoupled weight decay regularization,''
\newblock in {\em Proc. ICLR}, 2019.

\end{thebibliography}
}

\end{document}